\documentclass[aps,twocolumn,showpacs,amsmath,amssymb,superscriptaddress,reprint]{revtex4-1}

\usepackage[utf8]{inputenc}
\usepackage{graphicx}
\usepackage[colorlinks=true,citecolor=blue]{hyperref}
\usepackage{units}

\usepackage{pdfpages}
\usepackage{pgffor}
\usepackage{etoolbox} 

\makeatletter
\patchcmd{\@outputpage@head}{\@ifx{\LS@rot\@undefined}{}{\LS@rot}}{}{}{}
\makeatother

\DeclareMathOperator{\T}{T}
\newcommand{\up}{\uparrow}
\newcommand{\down}{\downarrow}
\renewcommand{\vec}[1]{\mathbf{#1}}

\newcommand{\kBT}{k_\text{B}T}

\begin{document}
\title{Odd-frequency superconductivity revealed by thermopower}
\author{Sun-Yong Hwang}
\affiliation{Theoretische Physik, Universit\"at Duisburg-Essen and CENIDE, D-47048 Duisburg, Germany}
\author{Pablo Burset}
\affiliation{Department of Applied Physics, Aalto University, FIN-00076 Aalto, Finland}
\author{Björn Sothmann}
\affiliation{Theoretische Physik, Universit\"at Duisburg-Essen and CENIDE, D-47048 Duisburg, Germany}
\date{\today}

\begin{abstract}
Superconductivity is characterized by a nonvanishing superconducting pair amplitude. It has a definite symmetry in spin, momentum, and frequency (time). While the spin and momentum symmetry have been probed experimentally for different classes of superconductivity, the odd-frequency nature of certain superconducting correlations has not been demonstrated yet in a direct way. Here we propose the thermopower as an unambiguous way to assess odd-frequency superconductivity. 
This is possible since the thermoelectric coefficient given by Andreev-like processes is only finite in the presence of odd-frequency superconductivity.
We illustrate our general findings with a simple example of a superconductor-quantum dot-ferromagnet hybrid.
\end{abstract}

\maketitle
\paragraph*{Introduction.--}
Superconductivity has generated an enormous interest ever since its discovery more than 100 years ago. It arises from the condensation of electrons into Cooper pairs.
Theoretically, Cooper pairs are described in terms of a superconducting pair amplitude $\mathcal F$, i.e., an anomalous Green's function consisting of two electronic annihilation or creation operators. The pair amplitude is characterized by a definite symmetry in spin, momentum and time (frequency) under the exchange of the two electrons forming a Cooper pair. Since $\mathcal F$ has to be odd under such an exchange, this yields the Berezinskii classification of superconductivity~\cite{berezinskii_new_1974}:
(i) even-frequency spin-singlet pairing which includes the well-known conventional $s$-wave pairing in BCS superconductors~\cite{bardeen_theory_1957} but also $d$-wave pairing in high-$T_c$ superconductivity~\cite{tsuei_pairing_2000};
(ii) even-frequency spin-triplet pairing in a $p$-wave state occurring in superfluid helium~\cite{vollhardt_superfluid_1990}, Sr$_2$RuO$_4$~\cite{mackenzie_superconductivity_2003} and topological junctions hosting Majorana fermions~\cite{fu_superconducting_2008};
(iii) odd-frequency spin-triplet $s$-wave pairing which can be generated in disordered Fermi liquids~\cite{kirkpatrick_disorder-induced_1991} and diffusive superconductor-ferromagnet hybrids~\cite{bergeret_odd_2005}; and
(iv) odd-frequency spin-singlet $p$-wave pairing~\cite{balatsky_new_1992}.
Odd-frequency pairing ubiquitously appears in hybrid junctions~\cite{tanaka_theory_2007,tanaka_odd-frequency_2007}, with all four classes present if spin-rotation symmetry is broken~\cite{tanaka_symmetry_2011,eschrig_spin-polarized_2015,linder_odd-frequency_2017}. 

The spatial symmetry of the Cooper pair wave function can be probed with a scanning tunneling microscope~\cite{pan_imaging_2000}. The spin symmetry is accessible, e.g., via spin-dependent transport measurements~\cite{keizer_spin_2006,khaire_observation_2010,robinson_controlled_2010}. However, the unambiguous detection of odd-frequency pairing is still an outstanding open issue. This is closely related to the fact that the superconducting pair amplitude is not a quantum-mechanical observable. 
Indirect evidence stems from the measurement of supercurrents through dirty ferromagnets which must be carried by odd-frequency pairs as they are immune to both impurity scattering and pair-breaking by the exchange field~\cite{bergeret_long-range_2001}. Furthermore, the presence of odd-frequency pairing can be inferred from a paramagnetic Meissner effect~\cite{yokoyama_anomalous_2011,alidoust_meissner_2014,di_bernardo_intrinsic_2015} and zero-bias peaks in the density of states~\cite{di_bernardo_signature_2015,alidoust_zero-energy_2015,pal_spectroscopic_2017}. An alternative theoretical proposal suggests using Majorana fermions as a probe for odd-frequency pairing~\cite{kashuba_Majorana_2017} which relies however on the critical assumption that Majorana fermions themselves are of odd-frequency nature. 

In this Rapid Communication, we propose the thermopower as a way to directly access the odd-frequency nature of Cooper pairs. 
The main idea is that a finite thermopower from Andreev-like processes is possible only in the presence of odd-frequency pairing. This is a general result that applies to any superconducting system. We remark that a finite Andreev thermopower requires the simultaneous presence of both, even- and odd-frequency pairing which is, however, the generic case in proximity-induced unconventional superconductivity~\cite{tanaka_symmetry_2011,eschrig_spin-polarized_2015,linder_odd-frequency_2017}.
There has been a considerable interest in the thermopower of superconductor-ferromagnet hybrids~\cite{machon_nonlocal_2013,ozaeta_predicted_2014,machon_giant_2014,giazotto_very_2015,hwang_large_2016,hwang_hybrid_2016,linder_spin_2016} but their potential to detect odd-frequency superconductivity has not been pointed out yet.
\begin{figure}
	\includegraphics[width=\columnwidth]{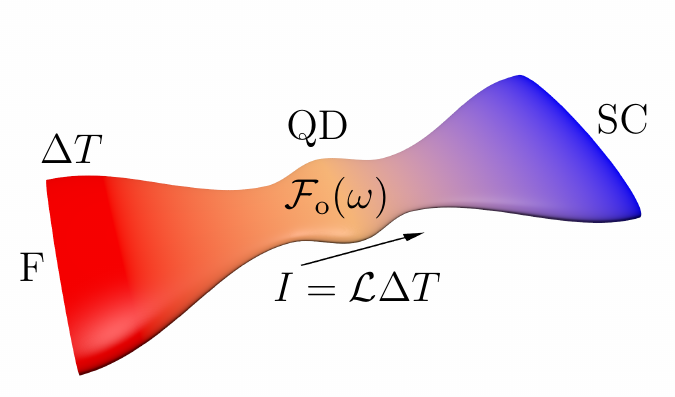}
	\caption{\label{fig:setup} Sketch of the setup. A mesoscopic conductor such as a quantum dot (QD) is connected to a BCS superconductor (SC) and a ferromagnet (F). Applying a temperature bias $\Delta T$ across the system gives rise to a thermoelectric charge current $I=\mathcal L\Delta T$.}
\end{figure} 
In a typical junction such as the one sketched in Fig.~\ref{fig:setup}, a temperature bias $\Delta T$ applied across the junction gives rise to a charge current which in linear response is given by $I=\mathcal L\Delta T$.
The thermoelectric coefficient $\mathcal L=S G$ is related to the thermopower (Seebeck coefficient) $S$ and the electric conductance $G$. Within a single-particle picture where interactions are treated on a Hartree-Fock mean-field level, it is given by~\cite{sivan_multichannel_1986,butcher_thermal_1990}
\begin{equation}\label{eq:L}
	\mathcal L=\frac{e}{hT}\int d\omega\; \omega \mathcal T(\omega)(-\partial_\omega f),
\end{equation}
where $-\partial_\omega f$ denotes the derivative of Fermi function and $\mathcal T(\omega)$ is the transmission probability of a quasiparticle with energy $\omega$ measured relative to the Fermi energy. Importantly, only the odd-in-frequency part of $\mathcal T(\omega)$ contributes to $\mathcal L$ since $\omega(-\partial_\omega f)$ is an odd function of $\omega$. This is in contrast to the electrical and thermal conductance which probe the even-in-frequency part of $\mathcal T(\omega)$~\cite{linder_quantum_2008}.
The transmission function can be related to the Green's function of the junction. 
By tuning the system to the particle-hole symmetric point (see below for details), the contribution from normal Green's functions can be strongly suppressed. The remaining contribution due to even- and odd-frequency anomalous Green's functions $\mathcal F(\omega)$ can be written as
\begin{equation}\label{eq:Teo}
\mathcal T(\omega)\propto |\mathcal F_\text{even}(\omega)+\mathcal F_\text{odd}(\omega)|^2.
\end{equation}
Hence, an odd-in-frequency part of $\mathcal T(\omega)$ exists if and only if odd-frequency pairing is present in the junction. Therefore, a finite thermopower is a smoking gun of odd-frequency Cooper pairs.

\paragraph*{Setup.--}
In the following, we are going to illustrate the general idea of our thermoelectric detection scheme with an example based on transport through a single-level quantum 
dot. 
It constitutes the simplest, yet experimentally relevant~\cite{hofstetter_ferromagnetic_2010} system that can exhibit both even- and odd-frequency pairing~\cite{weiss_odd-triplet_2017}. In addition, it offers a large degree of tunability where, in particular, odd-frequency pairing can be switched on and off via an external magnetic field.
Recently, thermoelectric effects have been studied experimentally in various quantum-dot setups~\cite{scheibner_thermopower_2005,scheibner_sequential_2007,svensson_lineshape_2012,svensson_nonlinear_2013,thierschmann_diffusion_2013,thierschmann_three-terminal_2015,dutta_thermal_2017}.

We consider a system consisting of a single-level quantum dot tunnel coupled to a ferromagnet and a conventional BCS superconductor.
The quantum dot has a single level with energy $\varepsilon_d$ and Zeeman splitting $B$. Double occupancy of the quantum dot with two electrons at the same time requires the Coulomb energy $U$. 
In the following, we first consider the limit of a noninteracting quantum dot, $U=0$ and discuss the influence of Coulomb interactions within a self-consistent Hartree-Fock approximation later on. 
The ferromagnetic lead is characterized by a spin-polarization $p=(\rho_+-\rho_-)/(\rho_++\rho_-)$ where $\rho_\pm$ denotes the density of states of majority and minority spin carriers at the Fermi energy. The ferromagnet is coupled to the dot level with spin-dependent coupling strength $\Gamma_{\text{F}\sigma}=(1\pm p)\Gamma_\text{F}$ and kept at a temperature $T+\Delta T$. 
The superconductor has an order parameter $\Delta$ which we choose to be real and positive without loss of generality. It couples with strength $\Gamma_\text{S}$ to the quantum dot and is kept at temperature $T$. 
The superconducting density of states in units of the normal-state density of states is given by $\rho_\text{S}=\Theta(|\omega|-\Delta)|\omega|/\sqrt{\omega^2-\Delta^2}$.

The coupling to the superconductor induces superconducting correlations on the quantum dot. They are characterized by the pair amplitudes $\mathcal F_{\sigma\sigma'}(t)=\langle \T d_{\sigma'}(t)d_\sigma(0) \rangle$ where $d_\sigma$ annihilates an electron with spin $\sigma$ on the dot. Since a single-level quantum dot does not have any spatial degrees of freedom, according to the Berezinskii classification only two types of superconductivity can be created: (i) even-frequency spin-singlet pairing $\mathcal F_e(t)$ and (ii) odd-frequency spin-triplet pairing $\boldsymbol{\mathcal F}_o(t)$~\cite{sothmann_unconventional_2014,burset_all-electrical_2016,weiss_odd-triplet_2017}. We thus parametrize the pair amplitude $F_{\sigma\sigma'}(t)=\left\{i\left[\mathcal F_e(t)+\boldsymbol{\mathcal F}_o(t)\cdot\boldsymbol{\sigma}\right]\sigma_y\right\}_{\sigma\sigma'}$, where $\boldsymbol{\sigma}$ denotes the vector of Pauli matrices.
The induced superconducting correlations are most conveniently characterized by defining order parameters~\cite{balatsky_even-_1993,abrahams_properties_1995,dahal_wave_2009,kashuba_Majorana_2017} which in thermal equilibrium at inverse temperature $\beta$ and to lowest order in the coupling to the ferromagnet are given by
\begin{align}
    \mathcal F_e(0)&=-\frac{i\pi\Gamma_\text{S}}{2\varepsilon_A}\frac{e^{-\beta\delta/2}\sinh \beta\varepsilon_A}{e^{-\beta\delta/2}\cosh\beta\varepsilon_A+e^{-\beta\varepsilon_d}\cosh\frac{\beta B}{2}},\\
    \partial_t \boldsymbol{\mathcal F}_o(0)&=\pi\Gamma_\text{S}\vec S-\frac{i}{2}\vec B\mathcal F_e(0),
\end{align}
where we defined $\delta=2\varepsilon_d+U$ and $2\varepsilon_A=\sqrt{\delta^2+\Gamma_\text{S}^2}$.
Hence, while coupling to a BCS superconductor always induces even-frequency correlations on the dot, odd-frequency correlations arise only if additionally a spin $\vec S$ accumulates on the dot or an external magnetic field $\vec B$ is applied.

\begin{figure*}
	\includegraphics[width=\textwidth]{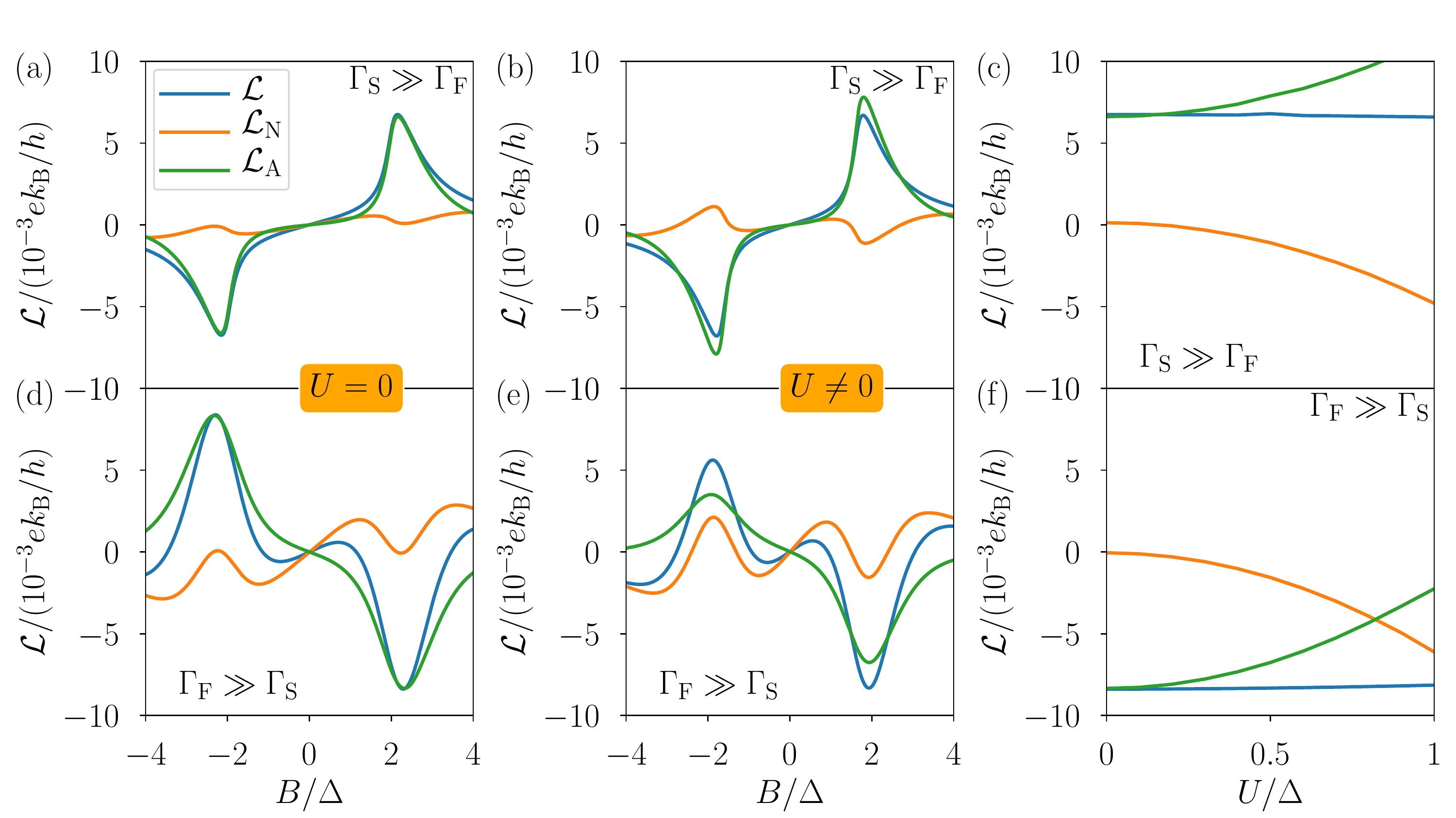}
	\caption{\label{fig:plots}(a) and (d) Thermoelectric coefficient $\mathcal L$ of the noninteracting quantum dot as a function of Zeeman splitting $B$. (b) and (e) Thermoelectric coefficient $\mathcal L$ as a function of Zeeman splitting $B$ for moderate Coulomb interactions $U=\Delta/2$. (c) and (f) Maximum of the thermoelectric coefficient as a function of Coulomb interaction $U$. In (a)-(c), the dot couples more strongly to the superconductor, $\Gamma_\text{S}=0.9\Delta$, $\Gamma_F=0.1\Delta$, while in (d)-(f) it couples more strongly to the ferromagnet, $\Gamma_\text{S}=0.1\Delta$, $\Gamma_F=0.9\Delta$. Other parameters are $\varepsilon_d=-U/2$, $p=0.3$, and $\kBT=0.2\Delta$.}
\end{figure*}

\paragraph*{Transmission function.--}
Since the only nonvanishing contribution of $\mathcal T(\omega)$ in Eq.~\eqref{eq:L} is given by its odd-in-frequency part, we henceforth redefine the transmission function  as its odd part only, $[\mathcal T(\omega)-\mathcal T(-\omega)]/2$.
A nonzero thermopower is a measure of the electron-hole asymmetry in the system. For a normal metal-quantum dot hybrid ($p=0$), this amounts to the condition $\varepsilon_d\neq -U/2$. In this case, the transmission function obtained via an equation-of-motion approach~\cite{cuevas_hamiltonian_1996,levy_yeyati_resonant_1997,burset_microscopic_2011} is given by~\cite{sup_mat}
\begin{equation}\label{eq:T0}
	\mathcal T(\omega)\propto \rho_\text{S}\Gamma_\text{S}\Gamma_\text{F}(2\varepsilon_d+U)\left[\omega+\frac{\Delta U}{2\omega}\left(\langle d_\up d_\down\rangle+\text{H.c.} \right)\right].
\end{equation}
Hence, $\mathcal L\propto2\varepsilon_d+U=0$ at $\varepsilon_d=-U/2$. Thus, by tuning the level position to the particle-hole symmetric point via a gate voltage, one is able to eliminate the undesired thermopower contribution which arises from a trivial breaking of particle-hole symmetry and is unrelated to odd-frequency pairing. This in turn allows for the unambiguous detection of odd-frequency pairing in the residual thermopower at $\varepsilon_d=-U/2$.

We now turn to the ferromagnet-quantum dot hybrid. For a noninteracting quantum dot at the particle-hole symmetric point, $\varepsilon_d=U=0$, the transmission function is given by
\begin{widetext}
\begin{equation}\label{eq:T}
	\mathcal T(\omega)=\rho_\text{S}p\Gamma_\text{F}\Gamma_\text{S}\sum_\sigma \left(|A_\sigma(\omega)|^2-|A_\sigma(-\omega)|^2\right)+\frac{\rho_\text{S}}{\rho_\text{S}^2-1}\frac{p\Gamma_\text{F}}{\Gamma_\text{S}}\left(\Gamma_\text{S}^2-\Gamma_{\text{F}\up}\Gamma_{\text{F}\down}\right)\left(|A_{\up\down}(\omega)|^2-|A_{\down\up}(-\omega)|^2\right),
\end{equation}
\end{widetext}
where $A_{\sigma}(\omega)$ and $A_{\sigma\bar\sigma}(\omega)$ are related to the ordinary and anomalous retarded Green's function of the dot, respectively (cf. Supplemental Material~[54]). We decompose Eq.~\eqref{eq:T} as $\mathcal T(\omega)=\mathcal T_\text{N}(\omega)+\mathcal T_\text{A}(\omega)$ and denote the corresponding contributions to Eq.~\eqref{eq:L} as $\mathcal L=\mathcal L_\text{N}+\mathcal L_\text{A}$. 
The first term, $\mathcal T_\text{N}(\omega)$, arises from remnant normal tunneling of unpaired quasiparticles in spin-asymmetric cases. 
The second term, $\mathcal T_\text{A}(\omega)$, is connected to odd-frequency pairing and originates from 
Andreev-like processes in which an electron (hole) from the ferromagnet with spin $\sigma$ picks up a quasiparticle (quasihole) of spin $\bar\sigma$ in the superconductor creating (annihilating) a Cooper pair~\cite{cuevas_hamiltonian_1996,sun_resonant_1999}. The term proportional to $\Gamma_{\text{F}\sigma}\Gamma_{\text{F}\bar\sigma}$ is reminiscent of Andreev reflection because of the retroreflection of a hole with an opposite spin.
These processes are mingled with the branch-crossing processes proportional to $\Gamma_\text{S}^2$ in Blonder-Tinkham-Klapwijk theory~\cite{blonder_transition_1982} where electrons (holes) incident from the ferromagnetic lead are converted into holelike (electronlike) excitations in the superconductor.
However, it should be noted that these processes appear at energies above the superconducting gap since there is no thermoelectric response in subgap transport~\cite{hwang_cross_2015}. 

\paragraph*{Thermopower.--}
According to Eq.~\eqref{eq:T} a finite thermoelectric response requires a finite polarization $p$. Furthermore, a finite thermopower occurs only in the presence of a finite Zeeman splitting $B$ which leads to a breaking of both particle-hole and spin symmetry. It is our main finding that for finite $p$ and $B$ one can maximize the thermopower contribution arising from odd-frequency pairing while minimizing or even eliminating completely the contribution from normal tunneling, $\mathcal L_\text{A}\gg\mathcal L_\text{N}$. This allows for an unambiguous detection of odd-frequency pairing in thermopower measurements.

A dominant thermopower contribution from odd-frequency pairing arises if either $\Gamma_\text{S}^2\gg\Gamma_F^2(1-p^2)$ or $\Gamma_\text{S}^2\ll\Gamma_F^2(1-p^2)$, i.e., for strongly asymmetric coupling [cf. Eq.~\eqref{eq:T}]. This is shown in Fig.~\ref{fig:plots}(a) and (d) where indeed $\mathcal L_\text{A}\gg\mathcal L_\text{N}$ in both cases. In particular, we find that the odd-frequency contribution $\mathcal L_\text{A}$ becomes maximal for Zeeman splittings equal to the superconducting gap, $B\approx 2\Delta$, while at the same time the normal contribution $\mathcal L_\text{N}$ vanishes. This allows for the interpretation of a finite thermopower as a smoking gun of odd-frequency pairing. We remark that the magnitude of the peak thermopower grows roughly linear with polarization as long as the asymmetric-coupling condition is fulfilled. In addition, the peak height is exponentially suppressed in $\Delta/T$ for low temperatures while it saturates around $T\sim\Delta/2$. 
For a realistic system based on an InSb nanowire quantum dot with an effective $g$-factor of $50$ and a Nb-based superconducting lead with $\Delta\approx\unit[1.5]{meV}$, a magnetic field of about $\unit[1]{T}$ is needed to achieve $B\approx 2\Delta$ which is well below the critical fields of Nb compounds.
Using in addition a ferromagnetic contact made from Fe, Co, or Ni with polarization $p=0.3$ and applying a temperature bias of $\unit[0.5]{K}$ yields a thermocurrent of about \unit[10]{pA}; readily measurable with current experimental technique.

\paragraph*{Coulomb interactions.--}
So far, we analyzed the case of a noninteracting quantum dot. While this allows for a transparent discussion of the underlying physics, Coulomb interactions play an important role in quantum dot physics. Therefore, we are now going to investigate how Coulomb interactions inside the quantum dot affect the results presented above. To address this question, we use the self-consistent Hartree-Fock approximation~\cite{anderson_localized_1961}. As before, we consider the particle-hole symmetric point $\varepsilon_d=-U/2$ where the contribution from Eq.~\eqref{eq:T0} vanishes. Within the Hartree-Fock approximation, the first term in Eq.~\eqref{eq:T}, $\mathcal T_\text{N}(\omega)$, remains unchanged if the definition of $A_\sigma(\omega)$ is generalized to take Coulomb interactions into account, see Supplemental Material~[54] for details. The second term arising from odd-frequency pairing takes the form
\begin{multline}\label{eq:Todd}
\mathcal T_\text{A}(\omega)=
\frac{\rho_S}{\rho_S^2-1+\left(\langle d_\up d_\down\rangle+\langle d_\down^\dag d_\up^\dag\rangle\right)^2\frac{U^2}{\Gamma_\text{S}^2}}\frac{p\Gamma_F}{\Gamma_\text{S}}\\
\left\{\Gamma_\text{S}^2-\Gamma_{F\up}\Gamma_{F\down}+U^2\left[\left(\langle d_\up d_\down\rangle+\langle d_\down^\dag d_\up^\dag\rangle\right)^2\right.\right.\\
\left.\left.-4\vphantom{\left[\left(\langle d_\up d_\down\rangle+\langle d_\down^\dag d_\up^\dag\rangle\right)^2\right.}\langle n_{d\up}\rangle\langle n_{d\down}\rangle+1\right]\right\}\left(|A_{\up\down}(\omega)|^2-|A_{\down\up}(-\omega)|^2\right),
\end{multline}
where $n_{d\sigma}=d_\sigma^\dagger d_\sigma$ and all averages $\langle\cdots\rangle$ have to be determined self-consistently. The resulting thermopower for an interacting quantum dot is shown in Fig.~\ref{fig:plots}(b) and (e).

For a strong coupling to the superconductor, $\Gamma_\text{S}^2\gg\Gamma_\text{F}^2(1-p^2)$, the inclusion of Coulomb interactions does not lead to any qualitative changes of the thermopower compared to the noninteracting case [cf. Fig.~\ref{fig:plots}(b)]. On a quantitative level, the position of the thermopower peaks is slightly shifted. Nevertheless, the maxima of the total thermopower $\mathcal L$ and the odd-frequency contribution $\mathcal L_\text{A}$ still occur at the same Zeeman splitting. Furthermore, the thermopower contribution from normal processes $\mathcal L_\text{N}$ no longer vanishes exactly at the peak position. Interestingly, the signs of $\mathcal L_\text{N}$ and $\mathcal L_\text{A}$ at the peak differ [see Fig.~\ref{fig:plots}(c)], which allows for an unambiguous detection of the odd-frequency contribution.

In the opposite coupling limit, $\Gamma_\text{S}^2\ll\Gamma_\text{F}^2(1-p^2)$, Coulomb interactions have a stronger impact on the thermopower [see Fig.~\ref{fig:plots}(e)]. In particular, the thermopower is no longer an odd function of Zeeman splitting, $\mathcal L(B)\neq -\mathcal L(-B)$. However, the general symmetry relation $\mathcal L(p,B)=\mathcal L(-p,-B)$ still holds. As can be seen in Fig.~\ref{fig:plots}(f) the thermopower peaks now involve contributions from both the odd-frequency and the normal part. In particular, for $U\gtrsim \Delta/2$, the $\mathcal L_\text{N}$ becomes a sizable fraction of $\mathcal L_\text{A}$, which makes it challenging to disentangle their respective contribution from a measurement of the total thermopower $\mathcal L$. Nevertheless, for moderate Coulomb interactions. the thermopower due to Andreev processes is rather dominant, thus allowing for a detection of odd-frequency pairing.

\paragraph*{Conclusions.--}
We proposed thermopower measurements as a tool to reveal the presence of odd-frequency superconductivity. 
The main idea is that the thermoelectric coefficient due to Andreev-like processes is only finite if odd-frequency pairing is present in the system.
We illustrated our general concept with a simple, yet experimentally relevant example of a ferromagnet-quantum-dot-superconductor hybrid. 
In perspective, the proposed detection scheme can also be used to establish the presence of odd-frequency pairing in exotic material classes and other, more complicated, hybrid structures.

\paragraph*{Acknowledgments.--}
We acknowledge fruitful discussions with Jürgen König, Rafael Sánchez, Yukio Tanaka and Stephan Weiß as well as financial support from the Ministry of Innovation NRW via the ``Programm zur Förderung der Rückkehr des hochqualifizierten Forschungsnachwuchses aus dem Ausland''.
P.B. acknowledges funding from the European Union's Horizon 2020 research and innovation programme under the Marie Sk\l odowska-Curie Grant Agreement No. 743884. 


%

\newpage


\section*{Supplementary material}

# Supplemental material for "Odd-frequency Superconductivity Revealed by Thermopower"

Sun-Yong Hwang,[1] Pablo Burset,[2] and Björn Sothmann[1]

[1]*Theoretische Physik, Universität Duisburg-Essen and CENIDE, D-47048 Duisburg, Germany*
[2]*Department of Applied Physics, Aalto University, FIN-00076 Aalto, Finland*

## GREEN'S FUNCTIONS

The Hamiltonian describing the ferromagnet-quantum dot-superconductor hybrid is given by $\mathcal{H} = \mathcal{H}_C + \mathcal{H}_D + \mathcal{H}_T$ with

$$\mathcal{H}_C = \sum_{\alpha=F/S,k,\sigma} \varepsilon_{\alpha k} c^\dagger_{\alpha k \sigma} c_{\alpha k \sigma} + \sum_{\alpha,k} \left( \Delta_\alpha c^\dagger_{\alpha k \uparrow} c^\dagger_{\alpha \bar{k} \downarrow} + \text{H.c} \right), \quad (1a)$$

$$\mathcal{H}_D = \sum_\sigma \varepsilon_{d\sigma} d^\dagger_\sigma d_\sigma + U n_{d\uparrow} n_{d\downarrow}, \quad (1b)$$

$$\mathcal{H}_T = \sum_{\alpha,k,\sigma} \left( t_{\alpha\sigma} c^\dagger_{\alpha k \sigma} d_\sigma + \text{H.c} \right), \quad (1c)$$

where $\Delta_F = 0$, $\Delta_S = \Delta$, $t_{\alpha\sigma}$ represents the tunneling amplitude to each lead, and $\sigma$ corresponds to the spin of the electron.

We introduce the Nambu spinor representation $\hat{d}_\sigma = (d_\sigma, d^\dagger_{\bar{\sigma}})^T$ and $\hat{c}_{\alpha k \sigma} = (c_{\alpha k \sigma}, c^\dagger_{\alpha \bar{k} \bar{\sigma}})^T$. In this basis, we write the time-ordered Green's function for the dot ($\mathcal{T}$ denoting the time-ordering operator)

$$\mathbf{G}^t_{d,d}(t,t') = -i \begin{pmatrix} \langle \mathcal{T} d_\sigma(t) d^\dagger_\sigma(t') \rangle & \langle \mathcal{T} d_\sigma(t) d_{\bar{\sigma}}(t') \rangle \\ \langle \mathcal{T} d^\dagger_{\bar{\sigma}}(t) d^\dagger_\sigma(t') \rangle & \langle \mathcal{T} d^\dagger_{\bar{\sigma}}(t) d_{\bar{\sigma}}(t') \rangle \end{pmatrix}, \quad (2)$$

and that for the dot-lead coupling

$$\mathbf{G}^t_{\alpha k,d}(t,t') = -i \begin{pmatrix} \langle \mathcal{T} c_{\alpha k \sigma}(t) d^\dagger_\sigma(t') \rangle & \langle \mathcal{T} c_{\alpha k \sigma}(t) d_{\bar{\sigma}}(t') \rangle \\ \langle \mathcal{T} c^\dagger_{\alpha \bar{k} \bar{\sigma}}(t) d^\dagger_\sigma(t') \rangle & \langle \mathcal{T} c^\dagger_{\alpha \bar{k} \bar{\sigma}}(t) d_{\bar{\sigma}}(t') \rangle \end{pmatrix}. \quad (3)$$

One can then calculate the time evolution of these matrix Green's functions [1–4]

$$\begin{pmatrix} i\hbar\partial_t - \varepsilon_{d\sigma} & 0 \\ 0 & i\hbar\partial_t + \varepsilon_{d\bar{\sigma}} \end{pmatrix} \mathbf{G}^t_{d,d}(t,t') = \hbar\delta(t-t') + \begin{pmatrix} U & 0 \\ 0 & -U \end{pmatrix} \mathbf{G}^t_{d,d;\text{HF}}(t,t') + \sum_{\alpha,k} \begin{pmatrix} t_{\alpha\sigma} & 0 \\ 0 & -t_{\alpha\bar{\sigma}} \end{pmatrix} \mathbf{G}^t_{\alpha k,d}(t,t'), \quad (4a)$$

$$\begin{pmatrix} i\hbar\partial_t - \varepsilon_{\alpha k} & -\Delta_\alpha \\ -\Delta_\alpha & i\hbar\partial_t + \varepsilon_{\alpha\bar{k}} \end{pmatrix} \mathbf{G}^t_{\alpha k,d}(t,t') = \begin{pmatrix} t_{\alpha\sigma} & 0 \\ 0 & -t_{\alpha\bar{\sigma}} \end{pmatrix} \mathbf{G}^t_{d,d}(t,t'), \quad (4b)$$

where we have introduced the Hartree-Fock Green's function in order to take into account the Coulomb interaction:

$$\mathbf{G}^t_{d,d;\text{HF}}(t,t') = -i \begin{pmatrix} \langle \mathcal{T} n_{d\bar{\sigma}} d_\sigma(t) d^\dagger_\sigma(t') \rangle & \langle \mathcal{T} n_{d\bar{\sigma}} d_\sigma(t) d_{\bar{\sigma}}(t') \rangle \\ \langle \mathcal{T} n_{d\sigma} d^\dagger_{\bar{\sigma}}(t) d^\dagger_\sigma(t') \rangle & \langle \mathcal{T} n_{d\sigma} d^\dagger_{\bar{\sigma}}(t) d_{\bar{\sigma}}(t') \rangle \end{pmatrix}. \quad (5)$$

We apply the Hartree-Fock decoupling scheme [5] to the matrix elements of $\mathbf{G}^t_{d,d;\text{HF}}(t,t')$ as

$$\langle \mathcal{T} n_{d\bar{\sigma}} d_\sigma(t) d^\dagger_\sigma(t') \rangle \approx \langle n_{d\bar{\sigma}} \rangle \langle \mathcal{T} d_\sigma(t) d^\dagger_\sigma(t') \rangle + \langle d_{\bar{\sigma}} d_\sigma \rangle \langle \mathcal{T} d^\dagger_{\bar{\sigma}}(t) d^\dagger_\sigma(t') \rangle, \quad (6a)$$

$$\langle \mathcal{T} n_{d\sigma} d^\dagger_{\bar{\sigma}}(t) d^\dagger_\sigma(t') \rangle \approx \langle n_{d\sigma} \rangle \langle \mathcal{T} d^\dagger_{\bar{\sigma}}(t) d^\dagger_\sigma(t') \rangle - \langle d^\dagger_\sigma d^\dagger_{\bar{\sigma}} \rangle \langle \mathcal{T} d_\sigma(t) d^\dagger_\sigma(t') \rangle. \quad (6b)$$

Thereby one can relate $\mathbf{G}^t_{d,d;\text{HF}}(t,t')$ to $\mathbf{G}^t_{d,d}(t,t')$ via the Hartree-Fock self-energy

$$\boldsymbol{\Sigma}_{\text{HF}} = U \begin{pmatrix} \langle n_{d\bar{\sigma}} \rangle & \langle d_{\bar{\sigma}} d_\sigma \rangle \\ \langle d^\dagger_\sigma d^\dagger_{\bar{\sigma}} \rangle & -\langle n_{d\sigma} \rangle \end{pmatrix}. \quad (7)$$

In addition, by solving Eqs. (4a) and (4b) for the dot Green's function one has for the self-energy due to the dot-lead coupling

$$\boldsymbol{\Sigma}_\alpha(t,t') = \sum_k \begin{pmatrix} t_{\alpha\sigma} & 0 \\ 0 & -t_{\alpha\bar{\sigma}} \end{pmatrix} \mathbf{g}^t_{\alpha k}(t,t') \begin{pmatrix} t_{\alpha\sigma} & 0 \\ 0 & -t_{\alpha\bar{\sigma}} \end{pmatrix}, \quad (8)$$

where $\mathbf{g}^t_{\alpha k}(t,t')$ is the isolated lead Green's function, i.e., the solution of the differential equation

$$\begin{pmatrix} i\hbar\partial_t - \varepsilon_{\alpha k} & -\Delta_\alpha \\ -\Delta_\alpha & i\hbar\partial_t + \varepsilon_{\alpha \bar{k}} \end{pmatrix} \mathbf{g}^t_{\alpha k}(t,t') = \hbar\delta(t-t'). \quad (9)$$

Following the same steps for the dot retarded Green's function ($\{,\}$ indicating the anti-commutator)

$$\mathbf{G}^r_{d,d}(t,t') = -i\Theta(t-t') \begin{pmatrix} \langle\{d_\sigma(t), d_\sigma^\dagger(t')\}\rangle & \langle\{d_\sigma(t), d_{\bar\sigma}(t')\}\rangle \\ \langle\{d_{\bar\sigma}^\dagger(t), d_\sigma^\dagger(t')\}\rangle & \langle\{d_{\bar\sigma}^\dagger(t), d_{\bar\sigma}(t')\}\rangle \end{pmatrix}, \quad (10)$$

and by Fourier transform, one finds in energy space

$$\mathbf{G}^r_{d,d}(\omega) = \left[\mathbf{g}^{r,-1}_d(\omega) - \mathbf{\Sigma}_{HF} - \mathbf{\Sigma}^r_F(\omega) - \mathbf{\Sigma}^r_S(\omega)\right]^{-1}, \quad (11)$$

where $\mathbf{g}^r_d(\omega)$ is the isolated noninteracting dot retarded Green's function

$$\mathbf{g}^r_d(\omega) = \begin{pmatrix} \omega - \varepsilon_{d\sigma} + i0^+ & 0 \\ 0 & \omega + \varepsilon_{d\bar\sigma} + i0^+ \end{pmatrix}^{-1}, \quad (12)$$

and the retarded self-energies due to couplings are explicitly given by

$$\mathbf{\Sigma}^r_F(\omega) = -\frac{i}{2}\begin{pmatrix} \Gamma_{F\sigma} & 0 \\ 0 & \Gamma_{F\bar\sigma} \end{pmatrix}, \quad (13a)$$

$$\mathbf{\Sigma}^r_S(\omega) = -\frac{i}{2}\begin{pmatrix} \Gamma_S S_d(\omega) & -\Gamma_S S_o(\omega) \\ -\Gamma_S S_o(\omega) & \Gamma_S S_d(\omega) \end{pmatrix}, \quad (13b)$$

with

$$S_d(\omega) = \frac{\Theta(|\omega|-\Delta)|\omega|}{\sqrt{\omega^2-\Delta^2}} + \frac{\Theta(\Delta-|\omega|)\omega}{i\sqrt{\Delta^2-\omega^2}}, \quad (14a)$$

$$S_o(\omega) = \frac{\Theta(|\omega|-\Delta)\mathrm{sgn}(\omega)\Delta}{\sqrt{\omega^2-\Delta^2}} + \frac{\Theta(\Delta-|\omega|)\Delta}{i\sqrt{\Delta^2-\omega^2}}. \quad (14b)$$

We have used the wide-band approximation $\Gamma_{\alpha\sigma} = 2\pi|t_{\alpha\sigma}|^2 \sum_k \delta(\omega - \varepsilon_{\alpha k})$ and finally we neglect the spin-dependence of the superconducting density of states, i.e., $\Gamma_{S\sigma} = \Gamma_S$. We label the matrix elements of Eq. (11) as

$$\mathbf{G}^r_{d,d}(\omega) = \begin{pmatrix} G^r_{ee}(\omega) & G^r_{eh}(\omega) \\ G^r_{he}(\omega) & G^r_{hh}(\omega) \end{pmatrix}. \quad (15)$$

This matrix determines the transport properties of the junction.

## TRANSMISSION

For the noninteracting case, the transmission function is given by Eq. (6) of the main text with

$$A_\sigma(\omega) = \left[\frac{(\omega - \mathrm{sgn}(\sigma)B/4)^2 + 2(\varepsilon_d/p)(\omega - \mathrm{sgn}(\sigma)B/4)}{(\omega + \varepsilon_d - \mathrm{sgn}(\sigma)B/4)^2 + (\Gamma_{F\bar\sigma} + \Gamma_S|\omega|/\sqrt{\omega^2-\Delta^2})^2/4}\right]^{1/2} G^r_{ee}(\omega), \quad (16)$$

$$A_{\sigma\bar\sigma}(\omega) = G^r_{eh}(\omega), \quad (17)$$

$$A_{\bar\sigma\sigma}(\omega) = G^r_{he}(\omega). \quad (18)$$

In the main text, we restrict ourselves to the particle-hole symmetric point $\varepsilon_d = -U/2 = 0$ at which the thermopower contribution unrelated to odd-frequency pairing vanishes.

Including the Coulomb interaction in the mean-field approximation [5], one can generally write for the anomalous Green's functions from Eq. (11)

$$A_{\uparrow\downarrow}(\omega) = \frac{1}{D(\omega)}\left[i\frac{\mathrm{sgn}(\omega)\Delta\Gamma_S}{2\sqrt{\omega^2-\Delta^2}} - U\langle d_\uparrow d_\downarrow\rangle\right], \quad (19a)$$

$$A_{\downarrow\uparrow}(\omega) = \frac{1}{D(\omega)}\left[i\frac{\text{sgn}(\omega)\Delta\Gamma_S}{2\sqrt{\omega^2-\Delta^2}} - U\langle d^\dagger_\downarrow d^\dagger_\uparrow\rangle\right], \quad (19b)$$

with

$$\begin{aligned}D(\omega) =& \left(\omega - \varepsilon_{d\uparrow} - U\langle n_{d\downarrow}\rangle + i\frac{\Gamma_{F\uparrow}}{2} + i\frac{\Gamma_S|\omega|}{2\sqrt{\omega^2-\Delta^2}}\right)\left(\omega + \varepsilon_{d\downarrow} + U\langle n_{d\uparrow}\rangle + i\frac{\Gamma_{F\downarrow}}{2} + i\frac{\Gamma_S|\omega|}{2\sqrt{\omega^2-\Delta^2}}\right) \\ &- \left(i\frac{\text{sgn}(\omega)\Delta\Gamma_S}{2\sqrt{\omega^2-\Delta^2}} - U\langle d_\uparrow d_\downarrow\rangle\right)\left(i\frac{\text{sgn}(\omega)\Delta\Gamma_S}{2\sqrt{\omega^2-\Delta^2}} - U\langle d^\dagger_\downarrow d^\dagger_\uparrow\rangle\right).\end{aligned} \quad (20)$$

As in the noninteracting case, we tune the dot level to the symmetric point $\varepsilon_d = -U/2 \neq 0$ in order to eliminate the undesired thermopower from the conventional electron-hole symmetry breaking. Thus, $\varepsilon_{d\sigma} = \varepsilon_d + \text{sgn}(\sigma)B/2 = [\text{sgn}(\sigma)B - U]/2$ for our chosen gate potential where $B$ is the Zeeman splitting of the quantum dot level. Finally, $A_\sigma(\omega)$ in Eq. (6) of the main article in the Hartree-Fock approximation under the condition $\varepsilon_d = -U/2$ can be generalized into

$$A_\sigma(\omega) = \left[\frac{(\omega - \text{sgn}(\sigma)B/4)^2 + U(\langle n_{d\sigma}\rangle - \langle n_{d\bar\sigma}\rangle)(\omega - \text{sgn}(\sigma)B/4 + \Delta U\langle d_\uparrow d_\downarrow\rangle/\omega)}{(\omega - U[1/2 - \langle n_{d\bar\sigma}\rangle] - \text{sgn}(\sigma)B/4)^2 + (\Gamma_{F\bar\sigma} + \Gamma_S|\omega|/\sqrt{\omega^2-\Delta^2})^2/4}\right]^{1/2} G^r_{ee}(\omega). \quad (21)$$

---



\end{document}